\documentclass[review]{elsarticle}

\usepackage{amsmath, amssymb, , graphicx}

\usepackage{epstopdf}

\begin{document}
\begin{frontmatter}

\title{The Geometric Field (Gravity) as an Electro-Chemical Potential in a Ginzburg-Landau Theory of Superconductivity}

\author{Victor Atanasov$^{*}$\footnote{$^{*}$ The corresponding author}}
\address{Department of Condensed Matter Physics, Sofia University, 5 boul. J. Bourchier, 1164 Sofia, Bulgaria}
\ead{vatanaso@phys.uni-sofia.bg}


\begin{abstract}

We extend the superconductor's free energy to include an interaction of the order parameter with the curvature of space-time. This interaction leads to geometry dependent coherence length and Ginzburg-Landau parameter which suggests that the curvature of space-time can change the superconductor's type. The curvature of space-time doesn't affect the ideal diamagnetism of the superconductor but acts as chemical potential. In a particular circumstance, the geometric field becomes order-parameter dependent, therefore the superconductor's order parameter dynamics affects the curvature of space-time and electrical or internal quantum mechanical energy can be channelled into the curvature of space-time. Experimental consequences are discussed.

\end{abstract}


\end{frontmatter}

\section{Introduction}

The predictions of  the General Theory of Relativity (GTR), which offered an understanding of gravitation in terms of geometric field, were confirmed in a series of spectacular experiments\cite{P&R}. However, the interaction of the geometric field with material media, the way in which the curvature of space-time affects the electronic properties of condensed matters systems on a micro- and macro- scopic  scale is to a large extent unknown and due to its ability to be experimentally verified of interest and importance since this venue can provide an exit from the predicament that gravitation remains an observational phenomenon and not an interaction we can control and artificially create and explore. The exit from the predicament lies in condensed matter systems that can affect the properties of space-time, that is they are coupled to the geometric field in a reversible way and obey laws of conservation of energy. Therefore, by changing their state, energy can be channelled into the geometric field. In this paper we extend the Ginzburg-Landau theory of superconductivity to include an interaction of the order parameter with the geometric field and explore its consequences.

The interplay of geometry and topology in condensed matter systems has been well explored in low-dimensional magnetic systems. In support of the statement in the present paper, we remind an interesting result, namely a magnetic system (multiple interacting solitons) on a cylinder with constant radius cannot reach an absolute minimum of its magnetic energy due to the interaction between individual excitations. However, provided the underlying elastic support (the cylinder) is allowed to change radius, an interaction between the magnetic system's degrees of freedom and the geometry of the underlying support is established. Its form is very similar to the term discussed here. As a result of this extension the combined magnetic and geometric (elastic support) system reaches absolute minimum by relaxing the extra magnetic energy into deformation of the underlying support. Therefore, it is not an unique statement that we make in this paper, namely the underlying geometry can change as a result of the interaction with a condensed matter system\cite{Rossen}.

Before we proceed, we would like to make an important distinction between two things: i.) the effects of the background geometry on the quantum condensate and ii.) the effects of the wave-function ($\psi-$field) on the gravitational field. The background geometry enters the dynamics of the quantum condensate in two places, one is the coupling term which action we are going to discuss and the second is the covariant derivative which contains the Christoffel symbols. Throughout the paper we are going to assume that $\nabla$ represents covariant differentiation and contains a remnant of the curved background geometry. However, due to the particular limiting cases we are going to consider, the presence of the Christoffel symbols in the covariant derivatives of the wave-function is not affecting the conclusions we are going to make. Indeed, one is tempted to assume that the background geometry is set on a macroscopic level by the distribution of matter as governed by GTR. The coupling of the matter  content to the geometric field on a macro scale, that is the Einstein's gravitational constant $\approx 1.9 \times 10^{-26}$ m kg$^{-1}$ is hopelessly small to expect any interaction of the $\psi-$field with the gravitational one. However, as discussed in the Appendix, in the quantum condensate acting like a material medium an energy conservation relation emerges. This relation is analogous in meaning to the GTR equation (a form of conservation of energy on a macro scale) and involves a term containing an expression for the energy of the geometric field on a micro scale. Therefore, one can exploit this energy conservation relation to channel energy/change the curvature of the background field at least on microscopic scale. Moreover, as discussed in the Appendix, the non-vanishing effects the wave-function of a quantum condensate can have on the gravitational field are associated with the similar action the Bohmian quantum potential and the geometric field have in the hydrodynamic form of the Schr\"odinger equation - they enter on an equal footing. As a result, we are led to believe that they have similar nature and therefore interact on a microscopic scale outside of what GTR governs. Ideas and experimental attempts on gravity-superconductor interactions are discussed in \cite{ref1}. A strong point in favour of the ideas discussed here is the existence of experimental attempts at gravitational field generation via electric discharge through a superconducting medium\cite{ref2}. However, the utilised structures are not characterised and the presence of intrinsic Josephson junctions is neither proven, nor deliberately sought. Therefore, we believe the paper would act as a stimulus, explanatory framework and experimental guidance for a future systematic attempt at verification.

The paper is organised as follows. Section \ref{sec.extension} is devoted to the extension of the Ginzburg-Landau theory of superconductivity in order to include the interaction with the geometric field. The extra term is justified in this section but motivated in the Appendix. Section \ref{sec. consequences} discusses the consequences of the extra term and leads to Section \ref{sec. experiment} where the experimentally relevant conditions to verify the quantum condensate - geometric field interaction are discussed. Section \ref{sec. conclusions} concludes the paper which ends with an Appendix.

\section{Extension to the Ginzburg-Landau theory}\label{sec.extension}

The macroscopic phenomenological theory of superconductivity supposes that the superconductor is simply a charged quantum condensate superfluid. Its free energy was postulated by Ginzburg and Landau\cite{GL}. There are two degrees of freedom: i.) the complex valued superconductor's order parameter field
\begin{eqnarray}\label{trigonometric form psi}
\psi(\vec{r})=\left|   \psi(\vec{r}) \right| e^{i \theta(\vec{r})}=\sqrt{n_s(\vec{r})} e^{i \theta(\vec{r})},
\end{eqnarray}
where $n_s=\left|   \psi(\vec{r}) \right|^2$ is the superfluid (charge) density and $\left|   \psi(\vec{r}) \right| =0$ denotes the destruction of the superconducting state; ii.) the vector potential field $\vec{A}(\vec{r})$ for the electromagnetic degrees of freedom.

The Ginzburg-Landau free energy has the form
\begin{eqnarray}
F_{net}\left( \psi,  \vec{A} \right)= \int d^3 x  \mathcal{F}_{GL},
\end{eqnarray}
where
\begin{eqnarray}\label{GL_density}
\mathcal{F}_{GL}=\mathcal{F}_{L}+ \mathcal{F}_{grad} + w_{mag}.
\end{eqnarray}
Here the first term
\begin{eqnarray}
\mathcal{F}_{L}=\alpha \left|   \psi(\vec{r}) \right|^2 + \frac{\beta}{2} \left|   \psi(\vec{r}) \right|^4
\end{eqnarray}
is the Landau term of mean field theory type and codes the free energy in a spatially homogenious state. The second term
$\mathcal{F}_{grad}= \frac{\hbar^2}{2 m^{\ast}} \left| \nabla \psi(\vec{r}) \right|^2$ is of gradient type and codes the kinetic energy of the condensate correctly only near the critical temperature ($T_c$) where $\left|\psi \right| \ll 1$ is small. In the case of a charged fluid (superconductor) the theory needs slight modification, that is the addition of magnetic field energy density $w_{mag}=\frac{|\vec{B}|^2}{2 \mu_0}$ and the gradient is replaced by its gauge invariant version
\begin{eqnarray}
\nonumber \mathcal{F}_{grad}&=& \frac{\hbar^2}{2 m^{\ast}} \left| \nabla_A \psi(\vec{r}) \right|^2\\
&=& \frac{\hbar^2}{2 m^{\ast}} \left| \left(  \nabla  -  \frac{i e^{\ast}}{\hbar m^{\ast}}  \vec{A}(\vec{r}) \right) \psi(\vec{r}) \right|^2
\end{eqnarray}
Provided the interaction between the electrons in the material is short-ranged as compared to the scale on which $\psi(\vec{r})$ varies, the different terms have the following origin: i.) gradient term is the kinetic energy; ii.) $\left|   \psi(\vec{r}) \right|^4$ term represents interaction and iii.) $\left|   \psi(\vec{r}) \right|^2$ term represents a combination of electrical and temperature-dependent chemical potential. In the superconductor the chemical potential controls the pairing ability of the fermions and as Lagrange multiplier reveals the number of (density of) bosons, that is Cooper pairs.

In the present paper we are going to extend the Ginzburg-Landau theory with an extra term in order to encode the properties of the quantum condensate in curved space-time and the interaction of its wavefunction with the geometric field (and vice-versa). The extra free energy density that needs to be included in the expression (\ref{GL_density}) is
\begin{eqnarray}\label{geom_density}
\mathcal{F}_{geom}=\gamma R^{(3d)} \left|   \psi(\vec{r}) \right|^2,
\end{eqnarray}
where $R^{(3d)}$ is the induced three-dimensional Ricci scalar curvature and $\gamma=-\hbar^2/24m^{\ast}.$ This extra term preserves the symmetries of Ginzburg-Landau theory as they are, therefore the theory
\begin{eqnarray}\label{GL extended}
F_{net}\left( \psi,  \vec{A} \right)= \int d^3 x \left( \mathcal{F}_{GL}+\mathcal{F}_{geom} \right) 
\end{eqnarray}
is invariant under the same gauge transformations.

The justification of the geometric term can be traced from the time-dependent extension of the Ginzburg-Landau theory
\begin{eqnarray}\label{time evolution}
i \hbar \frac{\partial \psi}{\partial t} = \frac{\delta F_{net}}{\delta \bar{\psi}},
\end{eqnarray}
where $\bar{\psi}$ denotes complex conjugation. This can be split into real and imaginary parts with the help of the trigonometric substitution (\ref{trigonometric form psi}):
\begin{eqnarray}
\hbar \frac{\partial \theta}{\partial t} &=& - \frac{\delta F_{net}}{\delta n_s},\\
\hbar \frac{\partial n_s}{\partial t} &=& \frac{\delta F_{net}}{\delta \theta}.
\end{eqnarray}
These last two equations resemble Hamilton's equations of motion. Therefore, $n_s$ and $\theta$ are canonically conjugate. 

Dirac proposed a phase observable operator $\hat{\theta},$ presumably canonically conjugate to the number of particles operator\cite{Dirac}, that is $\hat{n_s}$ (using a harmonic oscillator as a toy model the commutator is given by $\left[  \hat{n}_s , \hat{\theta} \right] = i \, \hat{1}$). Unfortunately, defining these operators in the  general case has long been an unresolved problem in quantum mechanics\cite{CCR}.

However, in the case of a quantum condensate,  particles share a macroscopic quantum state,  and the condensate exhibits manifestly classical properties. In the case of a Josephson junction between superconductors the phase
difference between the two condensates is measurable as well as its particle number expressed as charge density. Anderson used the uncertainty relation
\begin{eqnarray}
\delta n_s \delta \theta \geq \frac12
\end{eqnarray}
between phase and charge density in the semi-classical context of superconductors\cite{Anderson}. Thus as operators in the case of charged quantum condensates,  $\hat{n}_s$ and $\hat{\theta}$  are quantum mechanically conjugate\cite{Legget, exp}.  As a result, the Heisenberg uncertainty principle stated above applies and can be read in the following manner: as the quantum condensate is destroyed $n_s=0$ ($\delta n_s=0$) (specified exactly), its phase is indefinite $\delta \theta = \infty$ (cannot be specified). In the present paper we will show a relation between the phase of the quantum condensate and the curvature of the geometric field (that is gravity). Since there exists a state in which the phase cannot be specified, then in this state the geometry of space-time cannot be specified to being exactly flat. In this state one can channel energy into the geometric field and create curved space-time configuration.

Now, let us return to the justification of the geometric term (\ref{geom_density}) to the free energy density (\ref{GL extended}). We put (\ref{GL extended}) into (\ref{time evolution}) to obtain
\begin{eqnarray}\label{time evolution}
i \hbar \frac{\partial \psi}{\partial t} = -\frac{\hbar^2}{2m^{\ast}} \nabla_A^2 \psi + \gamma R^{(3d)} \psi + \alpha \psi + \beta |\psi|^2 \psi,
\end{eqnarray}
which in the $\psi \to 0$ limit should coincide with the Schr\"odinger equation in curved space-time\cite{Victor} (see the Appendix). This leads to
\begin{eqnarray}
\gamma = - \frac{\hbar^2}{24 m^{\ast}}.
\end{eqnarray}
Note the particular form of the geometric field energy density is not a limitation since for a product space (that is 3+1 splittable space-time) there is a relation between the induced onto a three-dimensional hyper-surface Ricci scalar curvature $R^{(3d)}$ and the complete four-dimensional Ricci scalar curvature\cite{Ficken}
\begin{eqnarray}\label{4d-to-3d}
R=\frac43 R^{(3d)},
\end{eqnarray}
 therefore
\begin{eqnarray}\label{geom_density 4d}
\mathcal{F}_{geom}=\frac 34 \gamma R \left|   \psi(\vec{r}) \right|^2.
\end{eqnarray}

\section{Consequences}\label{sec. consequences}

Now let us explore some of the consequences of the extended theory. First, we turn to the Ginzburg-Landau length scales. Suppose fields vary in space but remain static in time. There are two length scales that can be constructed from the theory: i.) the penetration depth $\lambda$ associated with the vector potential $\vec{A}$ and ii.) the coherence length $\xi$ associated with the order parameter properties. Their ratio $\kappa=\lambda/\xi$ determines the type of superconductor at hand: i.) Type I for which  $0 <\kappa < 1/\sqrt{2}$ and ii.) for Type II $\kappa > 1/\sqrt{2}.$ The coherence length sets the range over which superconducting order is affected by local perturbations (the alternative view is the size of the Cooper pair radius). $\xi$ also represents the length over which $\psi$ returns to the bulk value from some disturbed one. The bulk value being the solution to 
${\delta F_{net}}/{\delta \bar{\psi}}=0$ while the kinetic term being at minimum, that is $\nabla_A \psi=0:$
\begin{eqnarray}
\nonumber && \alpha + \gamma R^{(3d)} + \beta \left|  \psi_0 \right|^2=0 \\\label{eq:psi_0}
&& \psi_0^2 = \frac{|\alpha| + |\gamma| R^{(3d)}  }{\beta},
\end{eqnarray}
where $|\alpha|=-\alpha$ and $|\gamma|=-\gamma.$ Here we see that the geometric field affects the order parameter in a manner similar to the action of the electro-chemical potential set by $\alpha.$ Note at $R^{(3d)}=0$ (flat space) Ginzburg-Landau result is restored $\psi_0^2 = {|\alpha| }/{\beta}.$ However, one can see a peculiar property, namely at sufficiently negative curvature of space-time $R^{(3d)}=-{|\alpha| }/{|\gamma|}$ the order parameter vanishes and superconductivity is destroyed. 

In order to determine the coherence length  we Taylor expand the order parameter around $\psi=\psi_0 + \delta \psi$ to get
\begin{eqnarray}
\left[ -\frac{\hbar^2}{2m^{\ast}} \nabla^2  + |\alpha| + |\gamma| R^{(3d)} \right] \delta \psi = 0
\end{eqnarray}
which is solved by $\delta \psi \propto e^{-x/\xi},$ where the coherence length is given by
\begin{eqnarray}
\xi = \sqrt{ \frac{\hbar^2}{2m^{\ast}}  \frac{1}{|\alpha| + |\gamma| R^{(3d)}}  }.
\end{eqnarray}
We see that the geometric field affects the length at which a variation of the order parameter decays. Ginzburg-Landau result is restored as geometry flattens. However, we notice a peculiar coincidence at the value  $R^{(3d)}=-{|\alpha| }/{|\gamma|}:$ the order parameter vanishes $\psi_0 \to 0$ while the coherence length $\xi \to \infty.$

In case the curvature of space-time is small we can expand the coherence length in series while keeping the lowest order terms
\begin{eqnarray}
\xi \approx \xi_{flat} \left( 1-  \frac{|\gamma| R^{(3d)}}{2 |\alpha|} \right) = \xi_{flat} \left( 1 - \frac{R^{(3d)} \xi_{flat}^2}{24}  \right),
\end{eqnarray}
where $\xi_{flat}=\sqrt{ \frac{\hbar^2}{2m^{\ast} |\alpha|} }.$ The characteristic curvature for which superconductor-geometric field interaction applies is given by 
\begin{eqnarray}
R^{(3d)} <  \frac{24}{\xi_{flat}^2}.
\end{eqnarray}
We see that the coherence length on flat space sets the limit curvature of space-time for which the weak field theory exposed here has a better chance to work.

Note, the currents from the Ginzburg-Landau theory, the London equations and the penetration depth are unaffected by the geometric field as the variation of the free energy with respect to the vector potential and the phase of the order parameter are unchanged:
\begin{eqnarray}
\frac{\delta F_{net}}{\delta \vec{A}}=0 &=&\frac{n_s e^{\ast}}{c} \vec{v}_s - \frac{1}{2 \mu_0} \nabla \times \left( \nabla \times \vec{A} \right)\\
\frac{\delta F_{net}}{\delta \theta}= 0& =& - \nabla . \vec{J}
\end{eqnarray}
The extra geometric term doesn't affect the conservation of charge/probability current
\begin{eqnarray}
\frac{\partial n_s}{\partial t}=- \nabla . \vec{J},
\end{eqnarray}
where $\vec{J}$ is the supercurrent operator 
\begin{eqnarray}
\vec{J}= \frac{\hbar}{m^{\ast}} n_s  \left( \nabla \theta - \frac{e}{\hbar c} \vec{A}  \right).
\end{eqnarray}
The penetration depth $\lambda$ is unaffected by the geometry but remains a function of temperature since $n_s(T):$
\begin{eqnarray}
\lambda^2=\frac{m^{\ast} c^2 }{ 4 \pi n_s e^2}.
\end{eqnarray}
As a result the Meissner effect continues to hold in curved space-time.

The Ginzburg-Landau parameter is equal to
\begin{eqnarray}
\nonumber \kappa=\frac{\lambda}{\xi}&=&\frac{\lambda/\xi_{flat} }{\sqrt{1+\frac{|\gamma|R^{(3d)}}{|\alpha|} } }\\
&\approx& \kappa_{flat} \left( 1 - \frac{R^{(3d)}\xi^2_{flat} }{24}  \right),
\end{eqnarray}
where $\kappa_{flat}$ is the Ginzburg-Landau parameter on flat space-time.
Note, {\it the geometry of space-time can change the type of superconducting material} at hand and this constitutes a verifiable prediction of the theory that we will further elaborate.

Now, let us go back to the issue with the chemical potential $\mu.$ The chemical potential of the superconducting phase is defined as
the rate of change of the free energy with respect to the change in the number of  Cooper pairs added to the condensate, that is
\begin{eqnarray}
\hbar \frac{d \theta}{d t}=-\frac{\delta F_{net}}{\delta n_s}=-\mu,
\end{eqnarray}
therefore
\begin{eqnarray}
\mu=\alpha + \beta |\psi|^2 + \gamma R^{(3d)}.
\end{eqnarray}
Indeed, {\it the curvature of space-time manifests as electro-chemical potential.} Next, suppose we have an electric potential drop between two points in the superconductor, then the phase difference between these points will grow linearly with time (the phase difference $\nabla \psi$ will grow as well), which is equivalent to the London equations, namely the Cooper pair condensate accelerates as free charged particles $\vec{J}_s \propto \nabla \theta.$ If nothing else takes place, it will reach critical current and the superconducting state will vanish. Therefore, a DC voltage is  not consistent with the steady state. Next, let us assume a steady state, that is $|\psi|^2=|\psi_0|^2$ (see eq. \ref{eq:psi_0}), then
$\hbar {d \theta}/{d t}=0,$ for every $R^{(3d)} \neq 0.$ The curved space-time doesn't affect the superconductor's ability to achieve a steady state.

We may assume the superconductor is not at steady state (which can happen at distances smaller than the coherence length) but rather at $|\psi|^2 \neq |\psi_0|^2.$ Furthermore, we require ${d \theta}/{d t}=0,$ that is the new state is stationary, while not necessarily equilibrium since $\vec{J}_s \propto \nabla \theta \neq 0$ (similar to DC current subject to Ohm's law: a non-equilibrium steady state; in case of a superconductor this can be realised at a Josephson junction). The Ginzburg-Landau theory is not valid for certain types of junctions, like S-N-S junctions. In the case where it applies
\begin{eqnarray}
\alpha=e^{\ast} U=- \beta |\psi|^2 + |\gamma| R^{(3d)}.
\end{eqnarray}
holds where $U$ is the electric potential drop. The above condition can be interpreted in the following ways: i.) $U=0,$ then
\begin{eqnarray}
\delta R^{(3d)}=- \frac{2 \beta}{|\gamma|} |\psi| \; \delta |\psi|,
\end{eqnarray}
that is the change of the embedding curvature of space-time can be induced by a change in the state of the superconductor, in effect the internal quantum mechanical energy of the condensate can be channeled into the geometric field; ii.) $U \neq 0,$ potential difference can be developed across a Josephson junction
\begin{eqnarray}
\alpha=e^{\ast} U=- \beta |\psi|^2 + |\gamma| R^{(3d)}.
\end{eqnarray}
and suppose the superconducting state is in the process of being destroyed $|\psi| \to 0,$ then
\begin{eqnarray}
R^{(3d)}= \frac{24 m^{\ast} e^{\ast} }{\hbar^2} U.
\end{eqnarray}
We can channel electrical energy into the geometric field via the mediation of a quantum condensate. Indeed, there is a proposition made in the paper, namely i.) { \it the superconductor's order parameter dynamics affects the curvature of space-time} as opposed to the standard way to read these relations, that is ii.) the geometry of space-time affects the dynamics of the order parameter. It is up to the ultimate instance in physics - the experimental verification to check if i.) applies as discussed here.

Now, let us explore how can the superconducting material medium determine the geometry of space-time. One possibility being the following approach: a.) include the time-dependent part of the evolution of the order parameter into the free energy density
\begin{eqnarray}\label{eq:F}
\mathcal{F}=\mathcal{F}_{GL} -\frac{i \hbar}{2} \left( \bar{\psi} \partial_t \psi - \psi \partial_t \bar{\psi}  \right)=\textrm{const.}; 
\end{eqnarray}
b.) extend the integration volume with the inclusion of the temporal dimension to obtain an invariant energy functional
\begin{eqnarray}\label{eq:int_F}
\int \mathcal{F} \sqrt{-g} d^4x=\textrm{const.}, 
\end{eqnarray}
where $g$ is the determinant of the metric of the four-dimensional space-time and c.) vary with respect to the components $g_{\mu \nu}$ of the metric tensor to obtain equations of motion that depend on geometry and reveal how the order parameter may determine the geometry of space time. We make use of (\ref{4d-to-3d}) and (\ref{geom_density 4d}) in the variation procedure to obtain  rather sophisticated equations of motion which exploration is beyond the scope of the present paper. However, we will demonstrate an approximate expression:
\begin{eqnarray}
\frac34 |\gamma| |\psi|^2 G^{\mu \nu} \approx -\frac12 g^{\mu \nu} \left(\mathcal{F} - \mathcal{F}_{geom}\right)
\end{eqnarray}
which reveals that the Einstein tensor is related to the free energy density of the condensate and as $\psi \to 0$ we may expect strong geometric effect.

\begin{figure}[h]
\begin{center}
\includegraphics[scale=0.25]{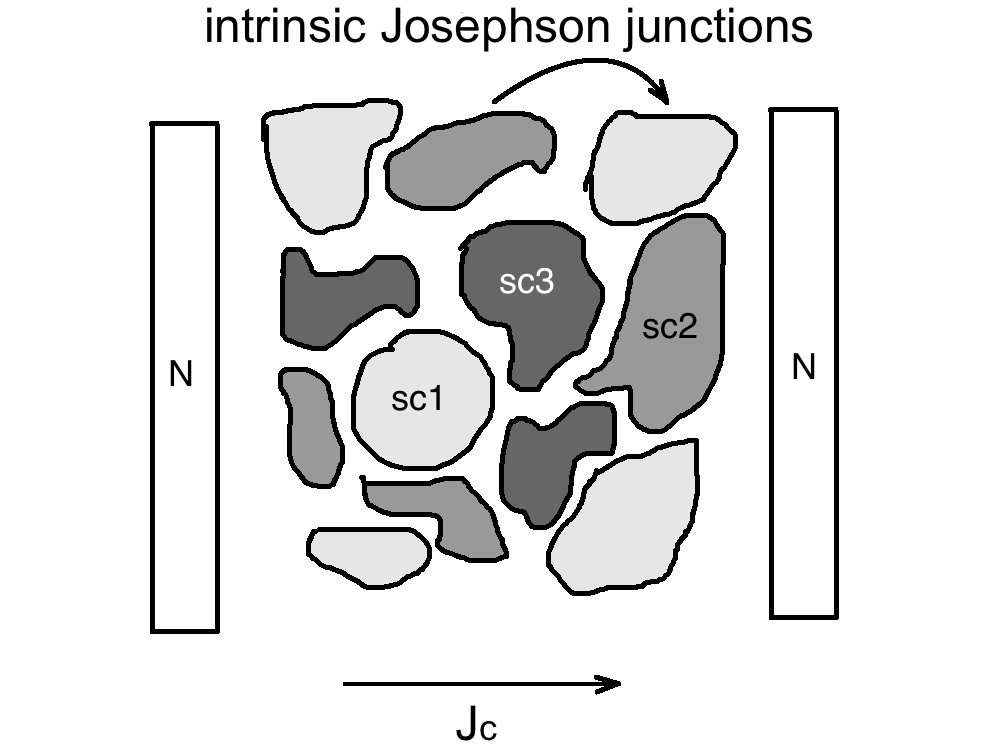}
\caption{\label{fig1} An electric discharge trough a granular superconducting structure made out of materials with different critical temperatures (sc1, sc2, sc3) and micro-tunnelling junctions between grains enacts both mechanisms in (\ref{eq:R, dtheta/dt, nabla psi}) to induce curvature in the geometric field by changing the state of the quantum condensate during the dischage. The contacts are made out of normal metal, denotes as N.}
\end{center}
\end{figure}

\section{Experimental verification}\label{sec. experiment}

Here, we will focus on a simplified but more straight-forward approach with a goal in mind: to propose an experimentally relevant context in which the effect can be confirmed. Let us take (\ref{eq:F}) and set the constant by exploiting the boundary condition of vanishing free energy as $\psi=0.$ This sets the constant to being equal the magnetic field density:
\begin{eqnarray}
\nonumber &&\qquad \lim_{\psi \to 0} \left(  \mathcal{F} =  \textrm{const.}\right) \quad \Rightarrow\\\label{eq:R, dtheta/dt, nabla psi}
&&\frac{1}{16} \; R =  \frac{2 m^{\ast}}{\hbar} \partial_t \theta  +  \frac{\left| \nabla_A \psi \right|^2}{|\psi|^2}; 
\end{eqnarray}
for a non-stationary (non-equilibrium) process the temporal change of phase is associated with an induction of space-time curvature (here $R$ is the four dimensional Ricci scalar). Alternatively, if we choose to read the relation in the other direction, if the quantum condensate is subject to a constant gravitational field its phase changes linearly in time: $\theta(t)= \frac{\hbar}{32 m^{\ast}} R (t-t_0).$ However, this doesn't necessary mean the condensate is accelerating, since acceleration requires $\nabla \theta \neq 0.$ Rather, the phase changes linearly with time at all points in space and the effect should not be noticed as is the experimental case of a quantum condensate (charged particle) in homogeneous gravitational field. A non-homogeneous gravitational field is required to produce acceleration in the quantum condensate.

We also notice that the non-vanishing gradient of the order parameter can also produce  curvature. Provided one creates conditions such that  $\nabla \psi \neq 0$ while $\psi \to 0,$ one can expect the strongest response of the geometric field. Note, the coupling between the change in the order parameter and the geometric field is direct in this particular process; {\it there aren't any coupling constants.} Therefore, we suggest the following experimental set up  which is characterised with a large probability to induce curvature of space-time by changing the state of the quantum condensate: suppose we have a granular superconducting material medium made out of several different superconducting materials. The size of the individual grains is larger than the coherence length for the particular material. They are mixed and tightly packed together in order for intrinsic Josephson junctions to form between individual grains (see Figure \ref{fig1}). These micro-tunnelling junctions are necessary in order to create intermittent structure for voltage drop to be build across individual grains and conditions to transfer electric energy into geometric field energy will be present. In the granular medium the super-current flows predominantly inhomogeneously $\nabla \psi \neq 0.$ Provided one conducts a voltage discharge through the structure, the value of the super-current will grow, that is $\partial_t \psi \neq 0,$ while reaching critical value and destroying the superconducting state, in effect $\psi \to 0$ and the two possible mechanisms to induce curvature encoded in (\ref{eq:R, dtheta/dt, nabla psi}) will be simultaneously exploited.

\section{Conclusions}\label{sec. conclusions}

In summary, we have extended the phenomenological Ginzburg-Landau theory of superconductivity to include an extra term providing the coupling between the geometric field (gravity) and the order parameter. As a result, the coherence length and the Ginzburg-Landau parameter of the superconductor become curvature dependent. The geometry of space-time can affect the superconductor by changing its type: from Type I to Type II and vice versa. However, the curvature of space-time doesn't affect the superconductor's ability to achieve a steady state, conserve charge/probability current and exhibit the Meissner effect, that is ideal diamagnetism is preserved in curved space-time. The resulting equations for the order parameter reveal that  the superconductor's order parameter dynamics affects the curvature of space-time in a manner which produces strongest response while the quantum condensate is being destroyed. This ability of the superconductor to induce curvature of space time is apt for experimental verification. A set up to test the effect is discussed in hope the present study invites experimental verification. 

The author would like to acknowledge the referee's valuable and constructive input.

\section*{APPENDIX}

The non-relativistic quantum mechanics in a 3d hyper-plane of the 4d curved space-time equipped with a metric can be treated in the following manner. Suppose the 4d space-time is splittable into 3+1 dimensions and the induced Riemannian metric $g_{ij}$ onto the 3d hyper-plane can be used to write the Laplace-Beltrami operator $\Delta_{LB}$, which is the kinetic energy term in the Schr\"odinger equation $
\Delta_{LB} \Psi = \frac{1}{\sqrt{|g|}} \partial_{j} \left( \sqrt{|g|} g^{jk} \partial_{k} \Psi  \right).$ The emergence of the geometric field from the kinetic term can be made clearer in the vicinity of the origin where the following Taylor expansion of the induced metric in normal coordinates applies: $g_{ij}=\delta_{ij}-\frac13 R_{ikjl}x^{k} x^{l} + O(|x|^3).$ Using a standard re-normalisation of the wave-function $\Psi=\psi/|g|^{1/4}$ and keeping the lowest order terms (the only relevant for the quantum dynamics) in the Taylor expansion we get\begin{eqnarray}
-\frac{\hbar^2}{2m}\Delta_{LB} \frac{\psi}{|g|^{1/4}} =  \frac{1}{|g|^{1/4}} \left( -\frac{\hbar^2}{2m}\Delta \psi - \frac{\hbar^2}{24m}R^{(3d)} \psi  \right) + O(|x|). 
\end{eqnarray}
Here $\Delta$ is the Laplacian on flat space. The emergence of a geometric potential is associated with the kinetic term. When electric field (defined with the potential $U$) and magnetic field, defined through the vector potential $\vec{A},$ are present the Schr\"odinger equation takes the following form
\begin{equation}\label{Schrodinger&R}
\frac{1}{2m}\left(\frac{\hbar}{i} \nabla - q \vec{A}  \right).\left(\frac{\hbar}{i} \nabla - q \vec{A}  \right) \psi + qU\psi+ \left( V_{Geom} + V \right) \psi=i\hbar \partial_t \psi, 
\end{equation}
where $V_{Geom}=- \frac{\hbar^2}{24m} R^{(3d)}.$ Here $R^{(3d)}$ is the three-dimensional Ricci scalar curvature. We may insert the trigonometric form of the wavefunction  $\psi=\sqrt{\rho(\vec{r})}e^{i\theta(\vec{r})},$ where $\rho(\vec{r})$ is the the charge density of the condensate and $\theta(\vec{r})$ its phase, in order to obtain the hydrodynamic form of the above equation. Taking the gradient of the real part of the equation (governing the dynamics of the phase) and expressing $\nabla \theta$ from the imaginary part (charge/probability conservation equation) we obtain a hydrodynamic interpretation of the quantum condensate dynamics:
\begin{eqnarray}\label{eq:dv/dt}
\frac{d \vec{v}}{dt}=\frac{1}{m} \vec{F}_{L} + \frac{1}{m}\nabla \left[ \frac{\hbar^2}{2m} \left( \frac{\Delta \sqrt{\rho}}{\sqrt{\rho}} + \alpha R^{(3d)} \right) \right], &&
\nabla \times \vec{v}=-\frac{q}{m}\vec{B}
\end{eqnarray}
where $\vec{F}_{L}=q\vec{E} + q \vec{v} \times \vec{B}$ is the Lorentz force acting on the charged Cooper pairs and $\alpha=1/12.$ These two equations are the equations of motion of the superconducting quantum fluid in the presence of space-time curvature. The geometric field enters the gradient of the Bohmian quantum mechanical potential, recognised by Bohm as an unique interaction with the $\psi-$field itself. The wave-function has the meaning of probability distribution. Probability is not derived from any material source. It is a pure information field on some stochastic process. Therefore, the geometric field and the $\psi-$field have identical action and probably meaning. 

Next, using the London equations (which hold true in curved space-time as well) for the quantum superconducting current $\vec{J}= -\rho q/m \vec{A},$ 
we obtain 
\begin{equation}
\vec{F}_{L} + \nabla \left[ \frac{\hbar^2}{2m} \left( \frac{\Delta \sqrt{\rho}}{\sqrt{\rho}} + \alpha R \right) \right]=-\frac{d }{d t} q \vec{A}.
\end{equation}

In the case of a robust superconducting state, we may assume that $\frac{d \rho}{d t} \approx 0$ and $\Delta \sqrt{\rho}\approx 0,$ then upon integration the following energy conservation relation applies in every material point:
\begin{eqnarray}
W_g(\vec{r}) - \mathcal{E}(\vec{r}) +  W_{int}(\vec{r}) ={\rm const}.
\end{eqnarray}
Here $\mathcal{E}(\vec{r})$ is the electrical energy of the Cooper pairs in the presence of an external electrical field, $W_{int}$ is the interaction energy of the Cooper pairs with the vector potential and  
\begin{equation}
W_g(\vec{r})=\frac{\hbar^2}{24m}  R^{(3d)}
\end{equation}
is the geometric field energy. This energy conservation relation implies the conversion of electric energy into geometric field energy, provided an electric field gradient can be built (in a Josephson junction this is translated into voltage-to-curvature conversion). 

Note, the existence of such an energy conservation relation represents an alternative to GTR mechanism to induce curvature of space-time. In order to stress the difference between the results these two frameworks predict, we convey a standard GTR argument, namely: suppose we include the gravitational field density from the Einstein-Hilbert action into (\ref{eq:int_F}) and keep the kinetic energy of the quantum condensate as the dominating term in the Ginzburg-Landau free energy density, then the two energy densities should be comparable provided
\begin{eqnarray}
\frac{c^4}{16\pi G} R \sim \frac{\hbar^2}{2 m^{\ast}} |\nabla \psi|^2 \approx \frac{\hbar^2}{2 m^{\ast} \xi^2 } |\psi|^2,
\end{eqnarray}
where $G$ is Newton's gravitational constant and $c$ the speed of light in vacuum. Here we have assumed that the coherence length $\xi$ is as small as necessary for $\nabla \psi \approx \psi/\xi$ to hold. The expected curvature of space-time should be of the order of
\begin{eqnarray}
R \sim \left( \frac{l_P}{\xi} \right)^2 \frac{8\pi \hbar}{c m^{\ast}} \;  |\psi|^2,
\end{eqnarray}
where $l_P$ is Planck's length. Therefore, according to GTR the induced curvature should be vanishingly small and unobservable, for any reasonable (including astrophysical) superconducting charge density. 

GTR governing equations are a form of conservation of energy relation acting on a macroscopic scale. Here, we have derived an energy conservation relation acting on a microscopic scale within a quantum condensate acting like a mediator. These two relations differ in form but not in meaning, therefore we expect experimental effort to either confirm or reject the proposed mechanism of geometric (gravitational) field generation in a condensed matter system.

\end{document}